\documentclass[a4paper]{jpconf}
\usepackage{graphicx}

\begin{document}

\title{Two models of spin glasses --- Ising versus Heisenberg
}
\author{Hikaru Kawamura}
\address{Department of Earth and Space Science, Faculty of Science,
Osaka University, Toyonaka 560-0043,
Japan}
\ead{kawamura@ess.sci.osaka-u.ac.jp}
\date{\today}
\begin{abstract}
Brief review is given on recent numerical research of the ordering of two typical models of spin glasses (SGs), the three-dimensional (3D) Ising SG and the 3D Heisenberg SG models. Particular attention is paid to the questions of whether there is a thermodynamic transition in zero field,  what are the associated critical properties,  what is the nature of the ordered state, particularly of a possible replica-symmetry breaking, and whether there is a thermodynamic transition in applied fields. The properties of the two models are contrasted, and possible relation to experiments is discussed.
\end{abstract}
%
%
%
%
%
%

\section{Introduction}

 Spin glasses are the type of random magnets in which both ferromagnetic and antiferromagnetic interactions coexist and compete, thereby giving rise to the effects of frustration and quenched randomness. Spin glasses (SGs) have been regarded as a typical example of complex systems, and their ordering properties  have been studied quite extensively, either experimentally, analytically or numerically \cite{review}.  Experimentally, we now have fairly convincing evidence that SG magnets exhibit an equilibrium  phase transition at a finite temperature into the glassy ordered state. This has been particularly well established in canonical SGs which represent a class of dilute transition-metal alloys soluted in the noble-metal host. 

 In theoretical studies of the ordering of SGs, a statistical model called the Edwards-Anderson (EA) model has been widely used \cite{EA}. In this model, spins are located on a certain regular lattice, say, on a three-dimensional (3D) simple cubic lattice, and are assumed to interact via the random exchange interaction taking either ferromagnetic or antiferromagnetic sign. One crucially important ingredient of the EA model is its spin symmetry. In its original version as proposed by Edwards and Anderson, the spin was assumed to be the Heisenberg one with three spin components. In fact, this is expected to be a good approximation for most of real SG materials, particularly for canonical SGs which possess relatively weak magnetic anisotropy. Another well-studied SG model is the Ising EA model with only one spin component. Although an ideal Ising SG is relative scarce in real SG materials, it has been widely used in numerous SG simulations performed in the past because of its simplicity. A naive expectation might be that the weak magnetic anisotropy inherent to real magnets eventually causes a crossover from the isotropic Heisenberg behavior to the anisotropic Ising behavior even in real SG materials \cite{Bray}, so that the Ising EA model corresponding to the strongly anisotropic limit is enough to describe the asymptotic behavior of weakly anisotropic SG magnets. Although such a naive expectation sounds plausible from the standard view of critical phenomena, the actual situation is not necessarily so simple as we shall see below.

 Obviously, it is also crucially important to fully understand the ordering properties of the Heisenberg EA model, in particular, whether the Heisenberg EA model behaves similarly or differently from the Ising EA model. Indeed, via recent extensive studies, it now becomes increasing clear that the 3D Heisenberg EA model exhibits an intriguing ordering behavior quite different from that of its Ising counterpart. Among others, a special degree of freedom called a ``chirality'', totally absent in the corresponding Ising model, might play a crucial role in the ordering process of the 3D Heisenberg EA model \cite{Kawamura10}, which might be crucially important in the ordering of real weakly anisotropic SG materials \cite{Kawamura10,Campbell10}.

 In the present article, we wish to review the recent studies on the ordering properties of the two typical models of SGs, {\it i.e.\/}, the 3D Ising EA model versus the 3D Heisenberg EA model, focusing on their similarities and differences. Indeed, although the recent numerical studies have indicated that both the Ising EA model and the Heisenberg EA model exhibit a finite-temperature transition, the nature of the ordering is often in sharp contrast with each other. In the present article, we wish to focus on the following four issues.

\noindent
(a) Is there a thermodynamic transition in zero field ?

\noindent
(b) If so, what are the associated critical properties ?

\noindent
(c) What is the nature of the ordered state ? Does it exhibit a replica-symmetry breaking (RSB) ? If yes, what type ?

\noindent
(d) Is there a thermodynamic transition in applied magnetic fields ?

 In the following, we shall examine the ordering properties of the 3D Ising SG in section 2, and those of the 3D Heisenberg SG in section 3. Section 4 is devoted to summary and discussion.

\section{The Ising spin glass}

 In this section, we review the ordering properties of the 3D Ising EA model. The Hamiltonian of the 3D Ising EA model is given by,
\begin{equation}
{\cal H}=-\sum_{<ij>} J_{ij} S_i S_j - H \sum_i S_i\ \ ,
\label{eqn:hamil}
\end{equation}
where $S_i=\pm 1$ is an Ising spin variable located  at the $i$-th site of a 3D simple-cubic lattice, $H$ is an external field intensity, and the $<ij>$ sum is taken over all nearest-neighbor pairs on the lattice. Periodic boundary conditions are applied in all directions. The nearest-neighbor couplings $J_{ij}$ are quenched random variables with zero mean and standard deviation unity. For the form of the $J_{ij}$ distribution, either the Gaussian distribution or the binary distribution ($\pm J$ distribution) is widely used. 

 The ordering properties of the 3D Ising EA model has been studied quite extensively for years as a standard model of SGs since the beginning of the SG research in 1970's. Note that the model has a finite-range (nearest-neighbor) interaction so that, in contrast to the mean-field-type (or an infinite-ranged) model such as the Sherrington-Kirkpatrick (SK) model \cite{SK}, the correlation effect is taken into account in the model. In spite of its apparent simplicity, however, the ordering properties of the 3D Ising SG model turns out to be highly nontrivial and hard to be elucidated. So far, rigorous analytical information on this model has been scarce, and most of the information we now have for this model is necessarily a numerical one. Nevertheless, after more than 30 years of extensive research, we now have several established (or at least almost established) knowledge about the ordering properties of the 3D Ising EA model. We summarize some of them below, together with several open issues.

\subsection{Ordering in zero field}

 In earlier studies, the issue of whether the 3D Ising EA model exhibits a finite-temperature transition or not remained controversial. Monte Carlo (MC) simulations on the model made by Ogielski and by Bhatt and Young in 1985 changed this situation, giving strong support to the occurrence of a finite-temperature SG transition \cite{Ogielski85,BhattYoung85}. Since then, many numerical works have been made, all of which have agreed in that the 3D EA model exhibits a thermodynamic SG transition at a finite temperature \cite{KawashimaYoung,Bernardi,Marinari98,Palassini99,MariCampbell99,Ballesteros00,MariCampbell02,Katzgraber06,Jorg06,Campbell06,Hasenbusch08}. In particular, the correlation-length ratio and the Binder ratio played an important role in identifying the SG transition point. In Fig.1, the correlation-length ratio $\xi_L/L$, defined by the finite-size SG correlation length $\xi_L$ divided by the linear size of the system $L$, calculated by Katzgraber {\it et al\/} for the $\pm J$ Ising EA model of $L\leq 32$ is shown, together with the Binder ratio calculated by the same authors \cite{Katzgraber06}. The correlation-length ratio and the Binder ratio are dimensionless quantities so that the data of various $L$ should be scale-invariant and exhibits a crossing behavior at the respective SG and chiral-glass transition points.  Indeed, clear fan-out of the correlation-length ratio and the Binder ratio is observed at $\beta^{-1}=T\simeq 1.12$, indicating the occurrence of a finite-temperature SG transition.

\begin{figure}[ht]
\begin{center}
\includegraphics[scale=0.9]{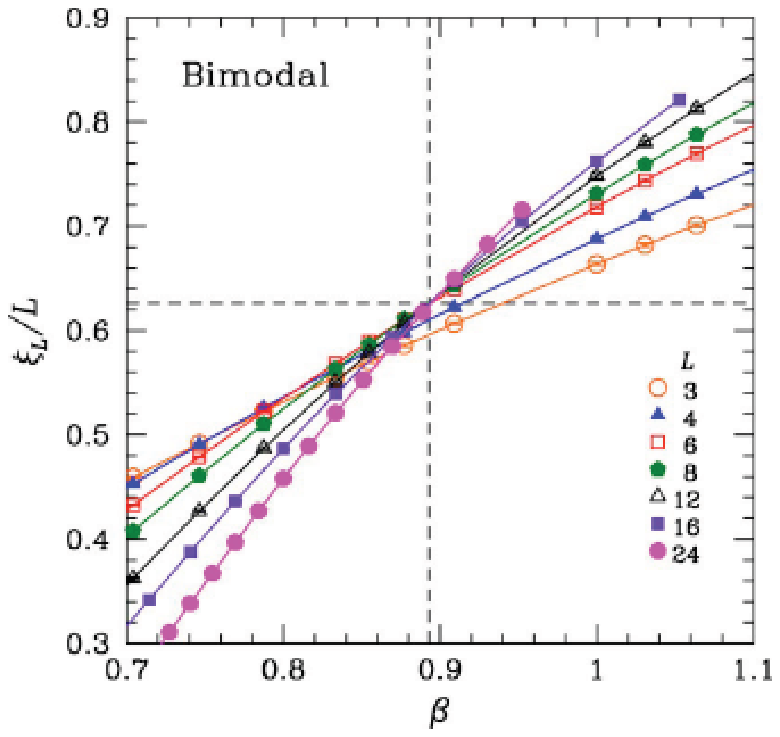}
\includegraphics[scale=0.9]{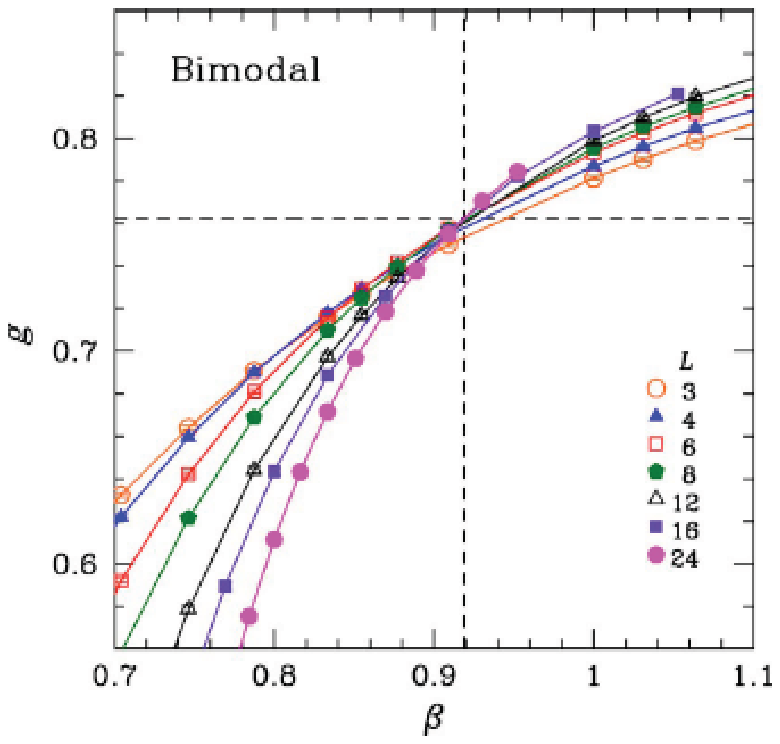}
\end{center}
\caption{
(Color online) The temperature and size dependence of the correlation-length ratio (upper figure), and the Binder ratio (lower figure), of the 3D Ising EA model with the binary ($\pm J$) coupling. $\beta$ is the inverse temperature. Taken from [H. G. Katzgraber, M. K$\ddot o$rner and A. P. Young, Phys. Rev. B{\bf 73} (2006) 224432].
}
\end{figure}

 By contrast, reliable estimates of the associated SG exponents are more difficult. Concerning the critical-point-decay exponent (or the anomalous dimension) $\eta$, various simulations have given nearly a common value of $\eta \simeq -0.38\sim -0.40$. By contrast, concerning the correlation-length exponent $\nu$, smaller values around $\nu\simeq 1.2-1.3$ reported by earlier simulations \cite{Ogielski85,BhattYoung85} have been revised by later simulations to significantly larger values. Indeed, most recent MC simulations have yielded $\nu \simeq 2.5\sim 2.7$, together with $\eta \simeq -0.38\sim -0.40$ \cite{Campbell06,Hasenbusch08}. The SG susceptibility exponent $\gamma$ is given by $\gamma \simeq 6.0\sim 6.5$. As an example, we show in Fig.2 the finite-size scaling plot of the correlation-length ratio of the $\pm J$ Ising EA model reported by Campbell {\it et al\/}, which yielded $\nu=2.72(8)$ \cite{Campbell06}.

\begin{figure}[ht]
\begin{center}
\includegraphics[scale=0.9]{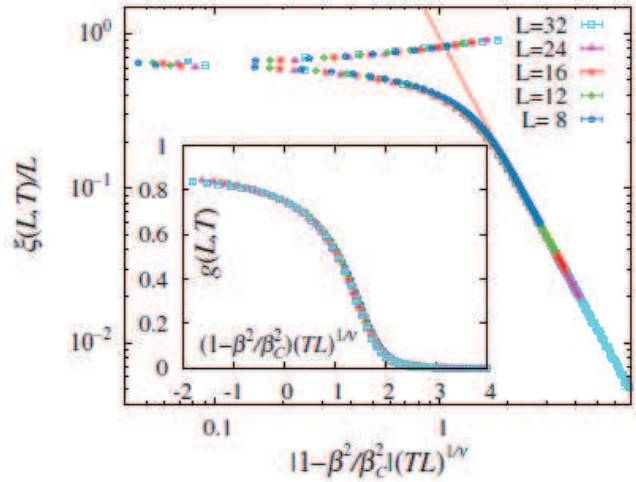}
\end{center}
\caption{
(Color online) Finite-size scaling plot of the correlation-length ratio of the 3D Ising EA model with the binary ($\pm J$) coupling. The exponent $\nu$ is taken to be 2.72. Taken from [I. A. Campbell, K. Hukushima, and H. Takayama: Phys. Rev. Lett. {\bf 97} (2006) 117202].
}
\end{figure}

 Although a good experimental realization of the Ising SG has been rather scarce, at least one example, an insulating SG magnet FeMnTiO$_3$, has been thoroughly studied. Indeed, the experimental exponent values measured for FeMnTiO$_3$ turns out to be roughly consistent with the numerical values of the 3D Ising SG \cite{Gunnarsson91,Campbell10}. 

 Meanwhile, in many of real SG magnets including canonical SGs, magnetic anisotropy is much weaker than the isotropic exchange interaction so that they might be described as a weakly anisotropic Heisenberg-like SG. It should be stressed that, in contrast to a naive expectation mentioned above, the experimentally determined SG critical exponents of such weakly anisotropic Heisenberg-like SGs largely deviate from the exponent values of the 3D Ising EA model. Indeed, for typical canonical SG materials, consistent experimental estimates are now available thanks to careful experimental measurements \cite{Simpson,Coutenary86,Bouchiat,Levy,Coles,Fert,Taniguchi88}. The exponents determined by various authors for canonical SGs indeed come close to each other, yielding the values $\beta \simeq 1$, $\gamma \simeq 2.2-2.3$, $\nu \simeq 1.3-1.4$ and $\eta \simeq 0.4-0.5$, which are clearly at odd with the exponent values of the 3D Ising EA model. The experimental $\nu $ and $\gamma$ are about half and the sign of $\eta $ is reversed.

 To elucidate the origin of the discrepancy observed between the critical properties of experimental Heisenberg-like SG magnets and of the 3D Ising SG is one of important issues left in the SG research.

\subsection{Possible replica-symmetry breaking}

 One of hot issues in the SG research has been concerned with the nature of the SG ordered state: In particular, whether the SG ordered state spontaneously breaks a replica symmetry or not \cite{review}. Two typical views have been common. One is a droplet picture, which claims that the SG ordered state is a ``disguised ferromagnet'' without a spontaneous RSB \cite{FisherHuse}. The other is a hierarchical RSB picture inspired by the exact solution of the mean-field model, which claims that the SG ordered state is intrinsically more complex accompanied with a hierarchical or full RSB where the phase space is hierarchically organized in the SG ordered state \cite{Parisi}. Hot debate has continued over years concerning which view applies to the ordered state of real SG magnets. 

 For more quantitative discussion, it is convenient to introduce an ``overlap'' variable $q$, which is defined by
\begin{equation}
q=\frac{1}{N}\sum_{i} S_i^{(\alpha)}S_i^{(\beta)}, 
\end{equation}
where $S_i^{(\alpha)}$ represents the $i$-th spin variable of the ``replica'' $\alpha$ and the summation is taken over all $N$ spins of the system. Replicas $\alpha$ and $\beta$ mean here the two independent copies of the system with the same realization of quenched randomness. One can then consider the distribution function of the overlap variable $P(q)$ 
\begin{equation}
P(q')=[<\delta(q-q')>], 
\end{equation}
where $<\cdots>$ represents a thermal average and [$\cdots$] an average over the quenched disorder (configurational average). Some typical forms of $P(q)$ in the thermodynamic limit is illustrated in Fig.3.

\begin{figure}[t]
\includegraphics[scale =0.45]{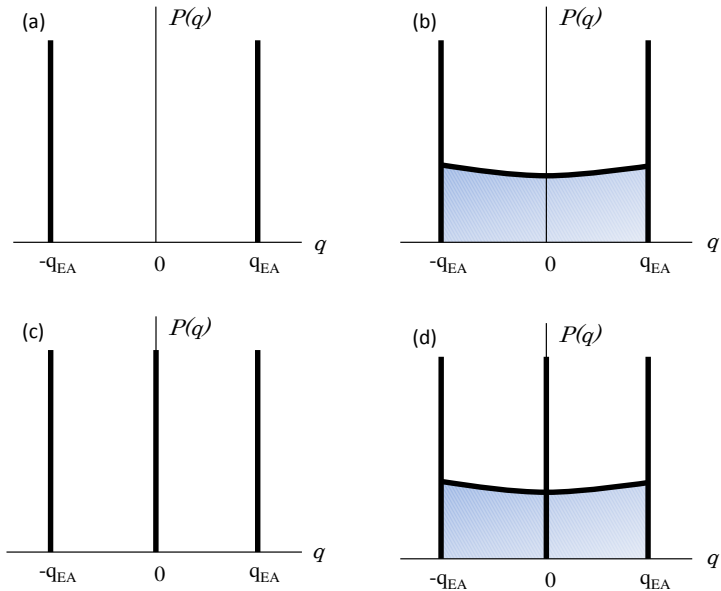}
\caption{
(Color online) Typical patterns of the overlap distribution function $P(q)$ in the thermodynamic limit. In the case (a) expected in a droplet theory, $P(q)$ consists of just two delta-function peaks at $q=\pm q_{EA}$. In the case of (b) expected in a hierarchical or full RSB picture, $P(q)$ possesses  an additional plateau part connecting the two delta-function peaks at $q=\pm q_{EA}$. In the case of (c) expected in a one-step RSB picture, $P(q)$ possesses  a central peak located at $q=0$ in addition to two delta-function peaks at $q=\pm q_{EA}$. The case (d) is a combination of (b) and (c).
}
\end{figure}

 The droplet picture claims that the overlap distribution describing the SG ordered state  to be a trivial one  in the thermodynamic limit consisting of just two delta functions located at $q=q_{EA}$ and at $q=-q_{EA}$. It means that the SG ordered state consists of unique pure state and its symmetry counterpart, irrespective of its apparent complexity in real-space spin pattern. 

 In the hierarchical RSB picture, by contrast, $P(q)$ exhibits a continuous plateau part spanning between the two delta-function peaks at $q=\pm q_{EA}$. It means that the phase space is divided into infinitely many pure states organized in a hierarchical manner, each of which is separated by infinitely high free-energy barrier. 
 
  Other types of RSB have also been known. One well-known example might be a one-step RSB, in which $P(q)$ possesses a central $\delta$-function peak at $q=0$ in addition to the self-overlap peaks  at $q=\pm q_{EA}$. In this case, the phase space is divided into many components, but all of them, except for itself and its symmetry partner, are completely dissimilar or orthogonal. It is realized, {\it e.g.\/}, in the ordered states of the mean-field Potts glass or of the mean-field $p$-spin model \cite{review,HukuKawaMF,Picco01}. In the past, such a one-step RSB has often been discussed in the context of the structural-glass problem rather than SG problem. The combination of the full RSB and the one-step RSB is also possible in certain models.

 Many numerical studies have focused on the 3D EA model to solve this issue of RSB, asking whether this model exhibits a hierarchical RSB or no RSB. The controversy has not yet been solved completely. However, most of the recent studies have indicated that the overlap distribution $P(q)$ persistently exhibits a nontrivial component suggesting the occurrence of a hierarchical-type RSB \cite{Marinari98,Katzgraber01,Katzgraber02}. As a typical example, we show in Fig.4  the overlap distribution $P(q)$ of the 3D Gaussian Ising EA model of various lattice sizes at a temperature $T=0.7$ which is well below the estimated SG transition temperature $T_g=0.95 \pm 0.04$ \cite{Marinari98}.

\begin{figure}[t]
\includegraphics[scale =0.95]{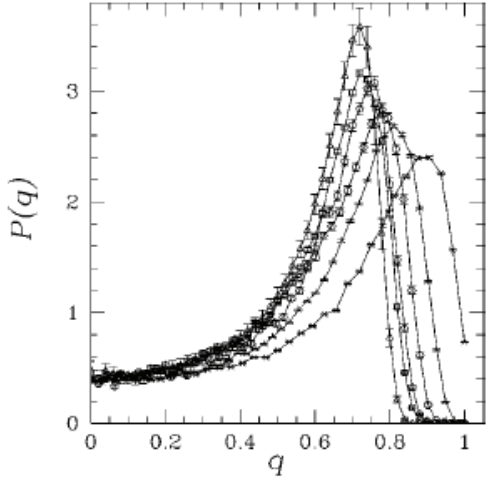}
\caption{
The overlap distribution function $P(q)$ in the spin-glass ordered state of the 3D Ising EA model with the Gaussian coupling. The temperature is $T=0.7$ where $T_g$ was estimated as $T_g=0.95\pm 0.04$. Only the positive side of $q$ is shown. The sizes are $L=4,6,8,10,12$ and 16. Taken from [E. Marinari, G. Parisi and J.J Ruiz-Lorenzo, Phys. Rev. B\textbf{58} (1998) 14852].
}
\end{figure}


\subsection{Ordering in finite fields}

 In contrast to the situation in zero field, the question of the existence/nonexistence of an equilibrium transition under external fields has remained quite controversial for years and not yet been completely settled. In fact, the question is closely related to the one discussed in the previous subsection, {\it i.e.\/}, that of an RSB. For example, if the SG ordered state of the 3D Ising EA model does not accompany an RSB, the model is likely to exhibit no finite-temperature transition under finite fields since applied fields destroy the global $Z_2$ spin-inversion symmetry, a symmetry broken at the zero-field transition. If, on the other hand, the SG ordered state of the model accompanies an RSB, a finite-temperature transition associated with RSB may become possible even in the situation where no global symmetry is left under fields, though it does not immediately necessitate the occurrence of an in-field transition.

 Although there is no completely consensus among researchers, most of recent numerical studies point to the absence of an equilibrium transition under finite fields \cite{Young04,Jorg08}. Such numerical observations seem consistent with the experimental observation on FeMnTiO$_3$: For this compound, Mattsson {\it et al\/} indicated that there was no in-field phase transition \cite{Nordblad95}, in sharp contrast to the zero-field case. Meanwhile, the absence of an equilibrium transition in finite fields seem in apparent contrast to the possible occurrence of a hierarchical-type RSB observed in zero field. Further study is required to clarify this point.

\section{The Heisenberg spin glass}

In the present section, we deal with the ordering of the 3D Heisenberg EA model. The Hamiltonian of the 3D Heisenberg EA model is given by,
\begin{equation}
{\cal H}=-\sum_{<ij>} J_{ij} \vec{S}_i \cdot \vec{S}_j - H \sum_i S_i^z \ \ ,
\label{eqn:hamil2}
\end{equation}
where $\vec{S}_i=(S_i^x,S_i^y,S_i^z)$ is now a three-component unit vector located  at the $i$-th site of a 3D simple-cubic lattice, $H$ is an external field intensity applied along the $z$-direction, and the $<ij>$ sum is taken over all nearest-neighbor pairs. Periodic boundary conditions are applied in all directions. The nearest-neighbor couplings $J_{ij}$ are quenched random variables with zero mean and standard deviation unity.  As in the Ising case, either the Gaussian distribution or the binary distribution ($\pm J$ distribution) is widely used for the form of the $J_{ij}$ distribution.

\subsection{Ordering in zero field}

 The ordering properties of the 3D Heisenberg EA model have long been studied \cite{Banavar,McMillan,OYS,Matsubara91,Yoshino93,Kawamura92,Kawamura95,Kawamura96,Kawamura98,HukuKawa00,Matsubara00,Endoh01,Matsubara01,KawaIma,Matsumoto,Nakamura02,LeeYoung03,BerthierYoung04,ImaKawa,Picco05,HukuKawa05,Campos06,LeeYoung07,Campbell07,Kawamura07,VietKawamura,Fernandez}, and are hotly debated even now. Earlier numerical simulations on the model suggested in common that the model exhibited only a zero-temperature transition \cite{Banavar,McMillan,OYS,Matsubara91,Yoshino93}, in apparent contrast to experiments. Common attitude in the community at that time was to invoke the weak magnetic anisotropy inherent to real materials to explain this apparent discrepancy with experiments, assuming that the weak anisotropy caused a rapid crossover from the $T_g=0$ Heisenberg behavior to the $T_g>0$ Ising behavior \cite{Bray}.  In fact, however, the situation was not quite satisfactory as already discussed above, in view of the fact that the experimental exponent values measured for canonical SGs are actually far from the Ising SG values, and no clear sign of Heisenberg-to-Ising crossover has been observed in experiments. 

 In 1992, the present author suggested that the 3D Heisenberg EA model might exhibit a finite-temperature transition {\it in its chiral sector\/} \cite{Kawamura92}. Chirality is a multispin variable representing the sense or the handedness of the noncollinear or noncoplanar structures induced by frustration, {\it i.e.\/}, whether the frustration-induced noncollinear or noncoplanar spin structure is right- or left-handed. The scalar chirality $\chi$ is defined by the product of three neighboring spins by 
\begin{equation}
\chi = \vec S_i\cdot  \vec S_j\times  \vec S_k.
\end{equation}
It has subsequently been suggested that, in the ordering of the 3D Heisenberg SG, the chirality was ``decoupled'' from the spin, the chiral-glass order taking place at a temperature higher than the SG order, $T_{CG} > T_{SG}$ \cite{Kawamura98,HukuKawa00,HukuKawa05,Kawamura07,VietKawamura}. Based on such a spin-chirality decoupling picture of the 3D isotropic Heisenberg SG, a chirality scenario of experimental SG transition was proposed by the author \cite{Kawamura92,Kawamura07}: According to this scenario, the chirality is a hidden order parameter of real SG transition. 

 The chirality scenario consists of the two parts, {\it i.e.\/}, the ``spin-chirality decoupling'' part for a completely isotropic system and the ``spin-chirality recoupling'' part for a weakly anisotropic system \cite{Kawamura10}. The first part, the spin-chirality decoupling, is a key ingredient of the scenario. It claims that the fully isotropic 3D Heisenberg SG exhibits a peculiar two-step ordering process, in which the systems exhibits, with decreasing the temperature, first the glass ordering of the chirality at a finite temperature $T=T_{CG}$ spontaneously breaking only a discrete $Z_2$ symmetry with preserving the continuous $SO(3)$ symmetry, and at a lower temperature $T=T_{SG}<T_{CG}$ exhibits the glass ordering of the spin itself fully breaking both the $Z_2$ and $SO(3)$ symmetries. The higher transition at $T=T_{CG}$ associated with the discrete $Z_2$ symmetry breaking is called the ``chiral-glass transition'', while the intermediate phase between $T_{CG}$ and $T_{SG}$ where only the $Z_2$ symmetry is broken with preserving the continuous $SO(3)$ symmetry is called the ``chiral-glass state''.  

 The second part of the chirality scenario concerns the role of the weak anisotropy which inevitably exists in real SG magnets to certain extent. The chirality scenario claims that the weak random magnetic anisotropy, which reduces the Hamiltonian symmetry from $Z_2\times SO(3)$ to only $Z_2$, weakly ``mixes'' the chirality to the spin sector, and the chiral-glass transition hidden in the chiral sector in fully isotropic system is `revealed' in the spin sector in weakly anisotropic system. In this scenario, the chiral-glass transition of the fully isotropic system, not the SG transition of the isotropic system, dictates the SG of real weakly anisotropic SGs.

 The chirality scenario is capable of explaining several long-standing puzzles concerning the experimental SG transition in a natural way, {\it e.g.\/}, the origin of the non-Ising critical exponents experimentally observed in canonical SGs, and remains to be an attractive hypothesis in consistently explaining various experimental observations for canonical SGs \cite{Kawamura10}. In recent numerical studies of the 3D Heisenberg EA model, although consensus now seems to appear that the 3D Heisenberg SG indeed exhibits a finite-temperature transition of some sort \cite{HukuKawa05,Kawamura07,Campos06,LeeYoung07,VietKawamura,Fernandez}, the nature of the transition, especially whether the model really exhibits the spin-chirality decoupling, is still under hot debate \cite{Matsubara00,Endoh01,Matsubara01,Nakamura02,LeeYoung03,BerthierYoung04,Picco05,Campos06,LeeYoung07,Fernandez}. 

 Thus, the issue of whether the spin-chirality decoupling really occurs in the 3D Heisenberg SG has remained controversial in spite of the potential importance in understanding the nature of experimental SG ordering. While several numerical results in favor of the occurrence of the spin-chirality decoupling were reported \cite{Kawamura98,HukuKawa00,KawaIma,Matsumoto,ImaKawa,HukuKawa05,VietKawamura,Fernandez}, a simultaneous spin and chirality transition without the spin-chirality decoupling was claimed in other works \cite{Matsubara00,Endoh01,Matsubara01,Nakamura02,LeeYoung03,BerthierYoung04,Picco05,Campos06,LeeYoung07,Fernandez}. The recent simulation of Ref.\cite{VietKawamura}, however, has provided a fairly strong numerical support for the occurrence of the spin-chirality decoupling. This calculation indicates that the SG transition point $T_{SG}$ is located about 10\% $\sim$ 15\% below the chiral-glass transition point $T_{CG}$, which we shall briefly describe below.

\begin{figure}[ht]
\begin{center}
\includegraphics[scale=0.9]{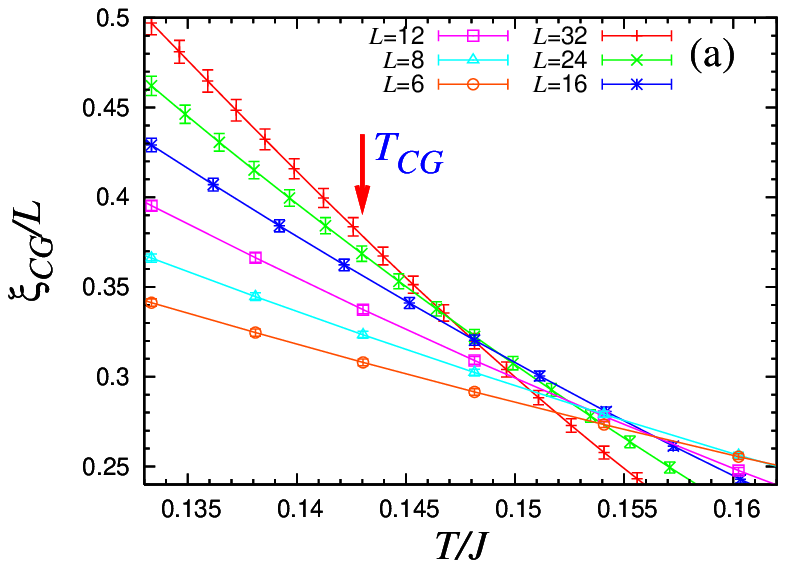}
\includegraphics[scale=0.9]{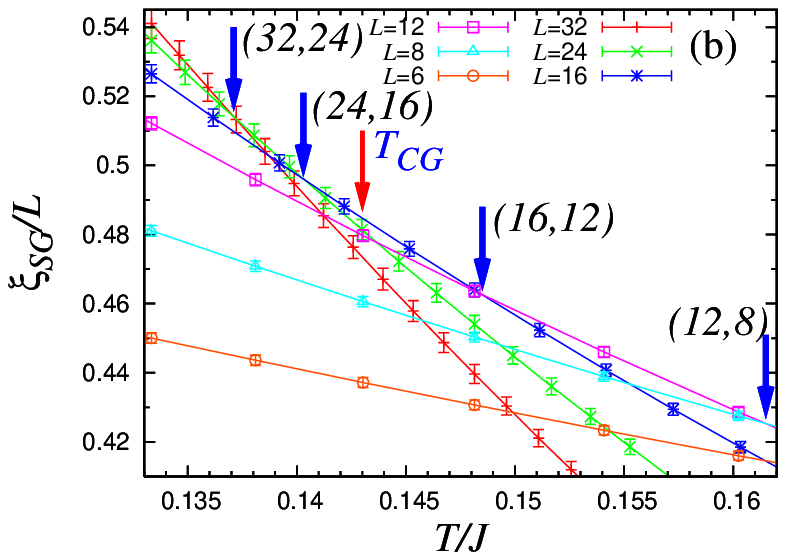}
\end{center}
\caption{
(Color online) The temperature and size dependence of the correlation-length ratio for the chirality (a), and for the spin (b). The arrow indicates the bulk chiral-glass transition point.
}
\end{figure}
\begin{figure}[ht]
\begin{center}
\includegraphics[scale=0.9]{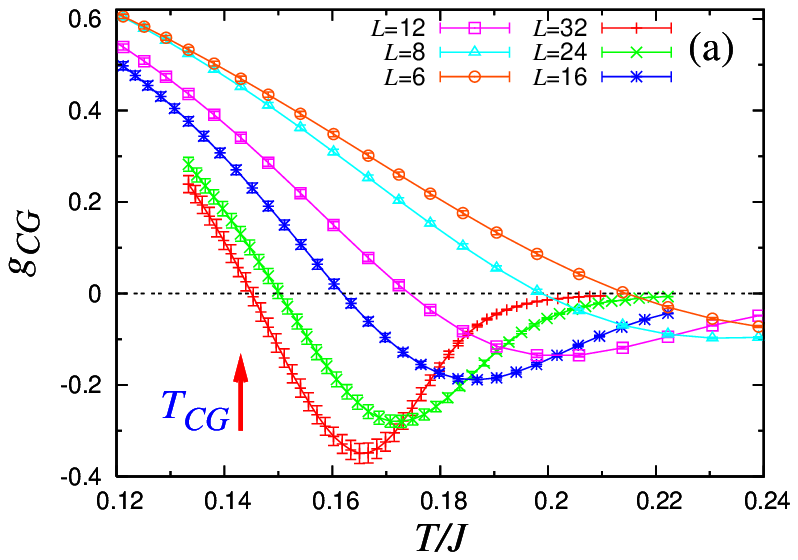}
\includegraphics[scale=0.9]{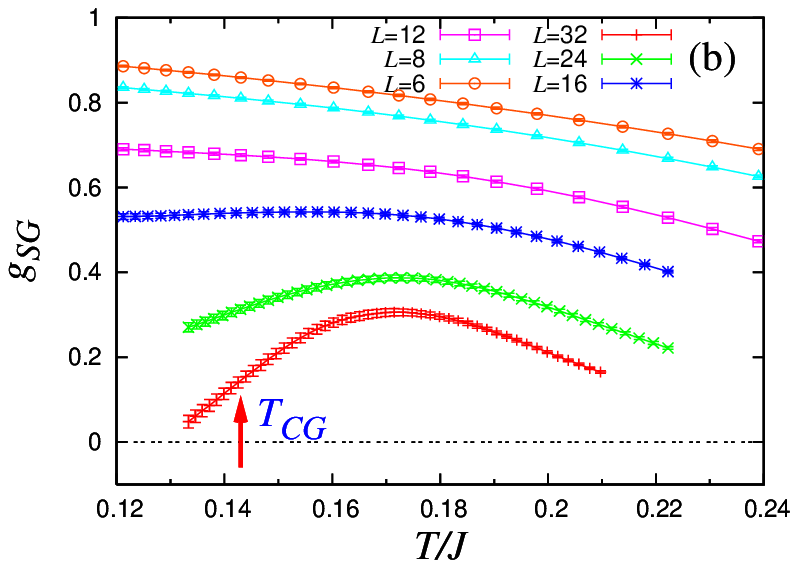}
\end{center}
\caption{
(Color online) The temperature and size dependence of the Binder ratio for the chirality (a), and for the spin (b). The arrow indicates the bulk chiral-glass transition point. 
}
\end{figure}
\begin{figure}[ht]
\begin{center}
\includegraphics[scale=0.9]{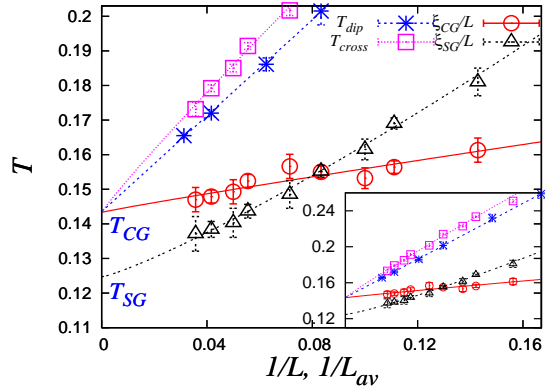}
\end{center}
\caption{
(Color online) The (inverse) size dependence of the crossing temperatures of $\xi_{CG}/L$ and  $\xi_{SG}/L$, the dip temperature $T_{dip}$ and the crossing temperature $T_{cross}$ of $g_{CG}$. The inset exhibits a wider range. 
}
\end{figure}
 In Fig.5, we show the recent MC data of the correlation-length ratios for the chirality  $\xi_{CG}/L$ (a), and for the spin  $\xi_{SG}/L$ (b), of the isotropic 3D Heisenberg EA model with random Gaussian coupling \cite{VietKawamura}. The system contains total $N=L^3$ spins with $L$ ranging from 6 to 32. For further details, refer to Ref.\cite{VietKawamura}. As can be seen from the figure, while the chiral $\xi_{CG}/L$ curves cross at temperatures which are only weakly $L$-dependent, the spin $\xi_{SG}/L$ curves cross at progressively lower temperatures as $L$ increases.  

 As an other indicator of the transition, we show in Fig.6 the Binder ratios for the chirality (a), and for the spin (b). The Binder ratios are also dimensionless, and are expected to exhibit a scale-invariant behavior at the respective chiral-glass and SG transition points. As can be seen from the figure, the chiral Binder ratio $g_{CG}$ exhibits a negative dip which deepens with increasing $L$. The data of different $L$ cross on the {\it negative\/} side of $g_{CG}$ unlike the correlation-length ratio. These features indicate a finite-temperature transition in the chiral sector. 

 To estimate the bulk chiral-glass and SG transition temperatures quantitatively, we plot in Fig.7 the crossing temperature of $\xi_{CG}/L$ and $\xi_{SG}/L$ for pairs of successive $L$ values versus $1/L_{av}$, where $L_{av}$ is a mean of the two sizes, together with  the dip temperature $T_{dip}$ and the crossing temperature $T_{cross}$ of the chiral Binder ratio  $g_{CG}$. The data show a near-linear $1/L_{av}$-dependence. The chiral crossing temperatures of  $\xi_{CG}/L$ and of $g_{CG}$ exhibits a weaker size dependence than the spin crossing temperature, and are extrapolated to $T_{CG}= 0.143\pm 0.004$. The spin crossing temperature exhibits a stronger size dependence, which is extrapolated to $T_{SG}= 0.125\pm 0.006$. Hence, $T_{SG}$ is lower than $T_{CG}$ by about 10\% $\sim$ 15\%.

\begin{figure}[t]
\includegraphics[scale =1.0]{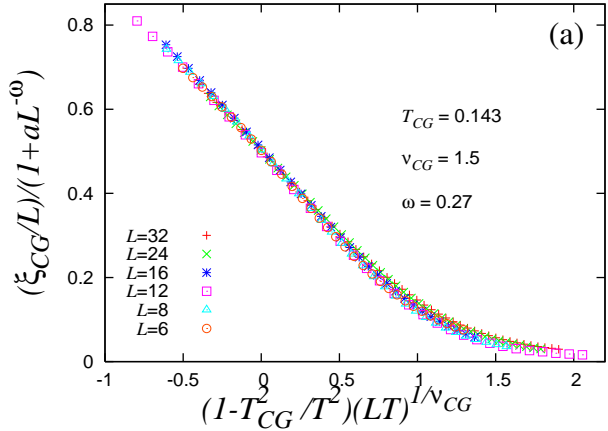}
\caption{
(Color online) Finite-size-scaling plot of the chiral-glass correlation-length ratio $\xi_{CG}/L$, where the correction-to-scaling effect is taken into account. The chiral-glass transition temperature and the leading correction-to-scaling exponents are  $T_{CG}=0.143$ and $\omega=0.27$. The best fit is obtained with $\nu_{CG}=1.5$.
}
\end{figure}

 Various chiral-glass exponents have been estimated via the standard finite-size scaling analysis. As an example, we show in Fig.8 the finite-size scaling plot of the chiral correlation-length ratio $\xi_{CG}/L$ in which the effect of the leading correction-to-scaling term has been taken into account. The exponents determined in this way from various physical quantities yield $\nu_{CG}= 1.5\pm 0.2$ and $\eta_{CG}= 0.6\pm 0.2$, {\it etc.\/}, which differ significantly from the standard 3D Ising SG values, $\nu \simeq 2.5\sim 2.7$ and $\eta \simeq -0.38\sim -0.40$ \cite{Campbell06,Hasenbusch08}. The results indicate that the chiral-glass transition belongs to a universality class different from that of the 3D Ising SG. Possible long-range and/or many-body nature of the chirality-chirality interaction might be the cause of this difference.

 Concerning the critical properties associated with the SG transition, which is likely to be located below the chiral-glass transition, the available information is still rather limited. The exponent $\eta_{SG}$ was reported to be $\eta_{SG}\simeq -0.3$ in Ref.\cite{VietKawamura}, while the exponent $\nu_{SG}$ to be $\nu_{SG}\simeq 1.5$ in Ref.\cite{Fernandez}.

\subsection{Possible replica-symmetry breaking}

 In this subsection, we discuss the issue of the nature of the chiral-glass ordered state, {\it i.e.\/}, whether the chiral-glass state exhibits an RSB, and if so, what type. In this connection, it should be noticed that the form of the chiral Binder ratio $g_{CG}$ shown in Fig.6, which exhibits a prominent negative dip, is quite peculiar. In fact, this form of the Binder ratio resembles the one of the system exhibiting a one-step RSB \cite{HukuKawaMF,Picco01}. 

 In Fig.9, we show the chiral-overlap distribution $P(q_\chi)$ in the chiral-glass phase  calculated in Ref.\cite{HukuKawa05} for the 3D Heisenberg SG model with the binary coupling. Here, the chiral overlap $q_\chi$ is defined by $q_\chi=\frac{1}{3N} \sum _{i\mu} \chi_{i\mu}^{(1)}\chi_{i\mu}^{(2)}$, where (1) and (2) indicate two copies (replicas) of the system. The local chirality variable $\chi_{i\mu}$ is defined at the $i$th site and in the $\mu$-direction ($\mu=x,y,z$) by $\chi_{i\mu}=\vec S_{i-e_\mu}\cdot (\vec S_i \times \vec S_{i+e_\mu})$ where $e_\mu$ is a unit lattice vector along the $\mu$-direction. The calculated $P(q_\chi)$ exhibits besides symmetric side peaks located at $q_\chi =\pm q_\chi^{{\rm EA}}$ corresponding to the long-range chiral-glass order, which grow with increasing $L$,  it also exhibits a prominent central peak at $q_\chi =0$, which also grows with increasing $L$. The existence of such a pronounced central peak is a characteristic feature of the system exhibiting a one-step-like RSB, never seen in the Ising SG. The data strongly suggest that the chiral-glass ordered state exhibits a one-step-like RSB \cite{HukuKawa00,HukuKawa05,VietKawamura}, in quite contrast to the one of the 3D Ising SG. Indeed, recent off-equilibrium experiments of the fluctuation-dissipation ratio of the Heisenberg SG magnet, thiospinel CdCr$_{1.7}$In$_{0.3}$S$_4$, seems consistent with such an expectation \cite{Kawamura10,HO}. Recent MC also indicates that the chiral-glass ordered state is non-self-averaging \cite{HukuKawa05,VietKawamura}.

\begin{figure}[t]
\includegraphics[scale =0.55]{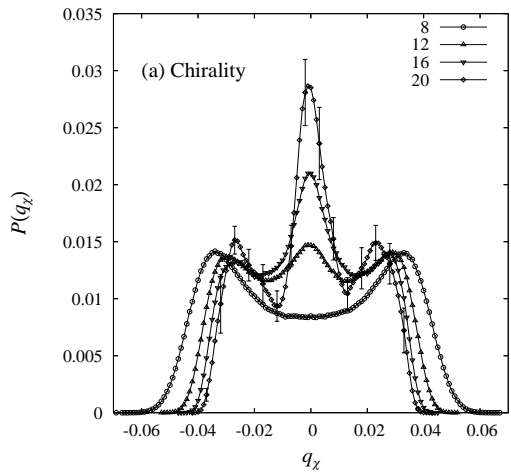}
\caption{
The overlap distribution function for the chirality of the 3D Heisenberg spin glass with the $\pm J$ binary coupling.  The temperature is $T=0.15$ below the chiral-glass transition temperature of this model $T=T_{CG}\simeq 0.19$.  [From K. Hukushima and H. Kawamura, Phys. Rev. B{\textbf 72} (2005), 144416.]
}
\end{figure}

\subsection{Ordering in finite fields}

 In this subsection, we discuss the ordering of the 3D Heisenberg SG under finite fields \cite{KawaIma}. When one applies a field to the fully isotropic Heisenberg SG, the symmetry of the Hamiltonian reduces from $Z_2\times SO(3)$ in zero field to $Z_2\times SO(2)$ under fields, where $Z_2$ refers to the chiral degeneracy associated with a spin-reflection operation (solely in spin space, not in real space) with respect to an arbitrary plane in spin space including the magnetic-field axis, while $SO(2)$ refers to the continuous degeneracy associated with a spin-rotation operation (in spin space, not in real space) around the magnetic-field axis in spin space. 

 Since the $Z_2$ chiral symmetry characterized by the sign of the scalar chirality remains under magnetic fields, the chiral-glass transition is expected to persist under magnetic fields. Of course, applied fields change the symmetry, but lower only the continuous part from $SO(3)$  to $SO(2)$. Since the continuous part has already been decoupled from the discrete $Z_2$ part, a natural expectation here would be that the chiral-glass transition persists even under fields essentially of the same type as the zero-field one. In particular, the chiral-glass transition line under fields should be a regular function of the filed intensity $H$. Since there is a trivial $H\leftrightarrow -H$ symmetry, the chiral-glass transition temperature under fields should behave for weak fields as $T_{CG}(H)\approx T_{CG}(0)- cH^2 \cdots $ ($c$ is a constant). In fact, this yields a transition line resembling the so-called GT line of the mean-field model \cite{GT}, $|T_{CG}(0)-T_{CG}(H)|\propto H^{1/2}$, although the origin of the exponent $1/2$ is entirely different: Here, 1/2 is just of regular origin, whereas the exponent $1/2$ in the mean-field model cannot be regarded as of regular origin. The expected phase diagram expected from the chirality scenario is sketched in Fig.10. 

\begin{figure}[t]
\includegraphics[scale =0.45]{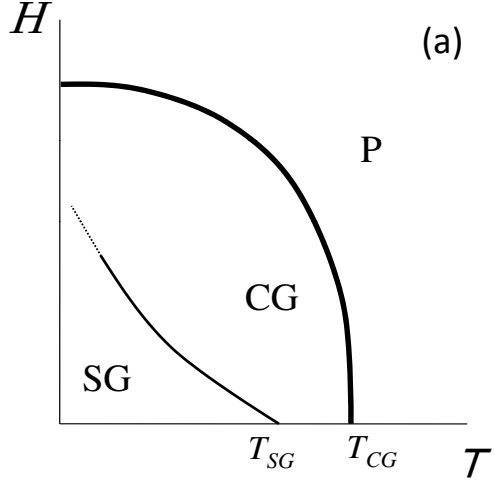}
\caption{
 Typical phase diagram of the 3D isotropic Heisenberg SG in the magnetic field ($H$) - temperature ($T$) plane, expected from the chirality scenario.  `P', `SG' and `CG' represent the paramagnetic, spin-glass and chiral-glass phases, respectively.
}
\end{figure}

 The ordering associated with the continuous part should also occur under finite fields. The in-field transition line associated with the continuous $SO(2)$ symmetry breaking should be a continuation of the $SO(3)$ breaking SG transition in zero field. Since the broken symmetry is different in zero and finite fields, {\it i.e.\/}, $SO(3)$ versus $SO(2)$, the SG transition line at low fields should exhibit a singular form, $H\propto |T_{SG}(0)-T_{SG}(H)|^{\phi/2}$. The exponent $\phi$ is not yet precisely determined, but may roughly be estimated as $\phi=\beta_{SG}+\gamma_{SG}\approx 4$. This $SO(2)$ breaking transition line is also sketched in Fig.10. 

 In-field ordering properties of the isotropic 3D Heisenberg SG were studied by Imagawa and the author also by means of MC simulations, which successfully demonstrates some of the feature of the phase diagram mentioned above \cite{KawaIma,ImaKawa}.  The obtained magnetic phase diagram is also consistent with the experimental one for canonical SGs \cite{Campbell10,Campbell99,Campbell02}.  In Fig.11, we show the magnetic phase diagram of the 3D isotropic Heisenberg EA model with binary coupling as determined by MC simulation \cite{KawaIma}. (Note that the SG transition line was not examined in this work).  Another interesting observation from MC is that the SG ordered state turns out to be quite robust against applied magnetic fields as can be seen from Fig.11 \cite{KawaIma,ImaKawa}. It appears to be stable up to fields as large as 25$k_B T_{SG}(H=0)$. This might be understandable if one notices that the coupling between the chirality and magnetic fields might be rather weak, since magnetic fields couple directly to the spin via the Zeeman term, only indirectly to the chirality.

\begin{figure}[t]
\includegraphics[scale =0.7]{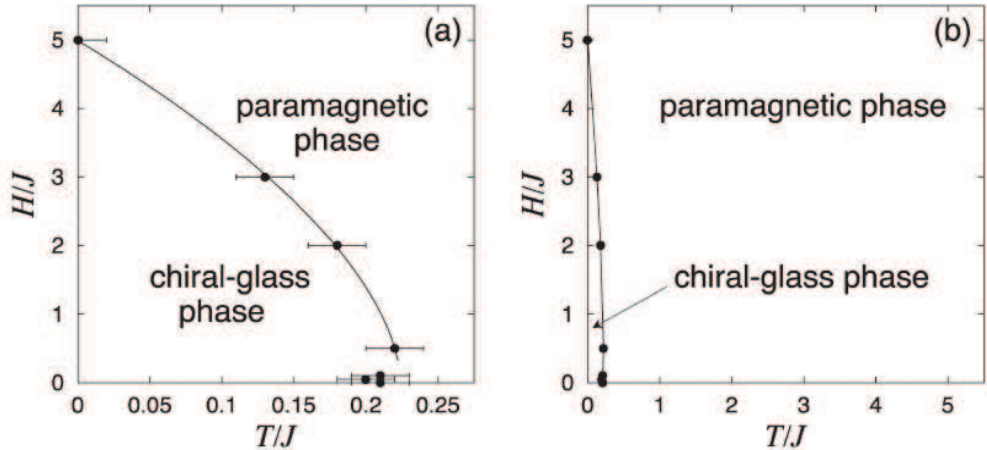}
\caption{
 Magnetic phase diagram of the 3D isotropic Heisenberg SG in the magnetic field ($H$) - temperature ($T$) plane. Note that the energy scales of the magnetic field $H$ and the temperature $T$ are mutually different in (a), while they are taken to be common in (b). Taken from [D. Imagawa and H. Kawamura, J. Phys. Soc. Jpn. {\bf 71} (2002) 127].
}
\end{figure}

\section{Summary and discussion}

 In this article, I reviewed the ordering properties of the two representative models of SGs, the 3D Ising EA model and the 3D Heisenberg EA model. Particular attention has been paid to i) whether the model exhibits a finite-temperature transition in zero field or not, and if it does, what are the critical properties of the transition; (ii) what is the nature of the ordered state, whether the model exhibits an RSB, and if yes, what type; (iii) whether the model exhibits a finite-temperature transition in finite fields.

 Concerning the zero-field ordering, both models, {\it i.e.\/}, the 3D Ising and Heisenberg EA models, exhibit a finite-temperature transition. A single SG transition occurs in the Ising EA model, whereas successive chiral-glass and SG transitions are likely to occur in the Heisenberg EA model, though there still remains controversy about the occurrence of such spin-chirality decoupling. Concerning the associated critical properties, recent numerical estimates yield $\nu_{SG}=2.5\sim 2.7$ and $\eta_{SG}=-0.38\sim -0.40$ for the 3D Ising SG, while, for the chiral-glass critical properties of the 3D Heisenberg SG, recent numerical estimates yield $\nu_{CG}\sim 1.5$ and $\eta_{CG}\sim 0.6$. For the SG exponents  of the 3D Heisenberg SG at its SG transition, which is likely to be located below the chiral-glass transition, recent numerical calculation suggests $\nu_{SG}\sim 1.5$ and $\eta_{SG}\sim -0.3$. One sees from these results that the universality classes associated with the spin ordering of the Ising SG, the chiral and the spin orderings of the isotropic Heisenberg SG are all different from each other. Numerically observed critical properties of the 3D Ising SG are roughly consistent with the experimental values of the Ising-like SG magnet FeTiO$_3$. Numerically observed chiral-glass critical properties of the 3D Heisenberg SG are close to those experimentally observed  for canonical SGs. This coincidence supports the chirality scenario of Refs.\cite{Kawamura92} which claims that the SG critical properties of real weakly anisotropic Heisenberg-like SG magnets such as canonical SGs should be equal to the chiral-glass ({\it not\/} SG !) critical properties of the fully isotropic Heisenberg SG.

  Concerning the issue of the RSB in the ordered state, numerical simulations on the 3D Ising SG point to the occurrence of a hierarchical-type RSB like the one in the infinite-range SK model, although the situation has remained somewhat controversial. For the 3D Heisenberg SG, recent simulations suggest that the ordered state might exhibit a one-step-like RSB, the one quite different from that of the Ising SG or of the mean-field SK model. According to the chirality scenario, such a one-step-like RSB should be realized in real weakly anisotropic SG magnets. 

 Concerning the in-field ordering properties, recent numerical simulations suggest that the 3D Ising EA model does not exhibit a finite-temperature transition in magnetic fields, which seems consistent with the experiment on the Ising-like  SG magnet FeMnTiO$_3$. By contrast, the 3D Heisenberg EA model exhibits a finite-temperature transition in magnetic fields associated with its chiral degrees of freedom, consistently with experimental observations for canonical SGs.

 Thus, the ordering properties of the 3D Ising SG and the Heisenberg SG are quite different from each other, each of which might provide a useful reference in attacking various problems related to SGs, not only of random and/or frustrated magnets but also of many systems including glassy materials, complex systems and even information problem.

\medskip

The author is thankful to I.A. Campbell,  D.X. Viet, K. Hukushima,  D. Imagawa for the collaboration and many useful discussion, and to T. Taniguchi, H. Yoshino, T. Okubo, M. Gingras, E. Vincent, M. Ocio, M. Picco and H. Takayama for discussion. This study was supported by Grant-in-Aid for Scientific Research on Priority Areas ``Novel States of Matter Induced by Frustration'' (19052006). We thank ISSP, Tokyo University and YITP, Kyoto University for providing us with the CPU time.

\end{document}